\documentclass[11pt]{article}
\usepackage{latexsym}  
\usepackage{amssymb}
\usepackage{graphicx}
\usepackage{amsmath}

\begin{document}

\begin{center}
{\Large\bf Neutrino Mass Matrix Related to  
Up-Quark Masses \\ and Nearly Tribimaximal Mixing \\[.1in]
 -- Based on a Yukawaon model --}\footnote{
A talk given at ``Particle Physics, Astrophysics and Quantum Field 
Theory: 75 Years since Solvay" (PAQFT08), 27-29 Nov. 2008,
Nanyang Executive Centre, Singapore.
To appear in the Conference Proceedings ({\em Intl.~J.~Mod.~Phys.~A}).
}\\[.1in]

{\bf Yoshio Koide}

{\it IHERP, Osaka University, Toyonaka, Osaka 560-0043, Japan} \\
{\it E-mail: koide@het.phys.sci.osaka-u.ac.jp}

\end{center}

\begin{abstract}
Based on a new approach (the so-called Yukawaon model) to the mass 
spectra and mixings,  a neutrino mass matrix which is 
described in terms of the up-quark masses and CKM matrix parameters 
is proposed. The mass matrix successfully leads to a nearly 
tribimaximal mixing without assuming any discrete symmetry.
\end{abstract}



\section{What is a Yukawaon model?}\label{sec1}
In the standard model of the quarks and leptons,
we have many parameters in the theory.
Especially, since the Yukawa coupling constants $Y_f$
 ($f=u,d,e,\nu$) are entirely free, we have no
 predictions for the mass spectra and mixings.
Therefore, usually, we assume a flavor symmetry, and
thereby, we discussed relations among the mass spectra
and mixings.
However, if we want to bring a flavor symmetry into the
standard model, we encounter a no-go theorem
\cite{no-go,no-go07} on the flavor symmetry, so that we cannot
impose any flavor symmetry on the standard model.
Of course, we can evade\cite{no-go07} the no-go theorem if we 
consider a multi-Higgs model in which Higgs scalars
have flavor quantum numbers.
However, such multi-Higgs models will newly bring
some troubles, e.g. flavor-changing neutral current problem,
rapid evolutions of coupling constants, and so on.

There is another idea for the origin of the mass spectra and mixings:
In the Yukawa interactions
$$
H_Y =\sum_{i,j}  \, \overline{q}_L^i (Y_u)_i^j u_{Rj} H_u
 + \cdots , 
\eqno(1.1)
$$
we regard the Yukawa coupling constants $Y_f$ as ``effective"
coupling constants $Y_f^{eff}$ in an effective theory, 
and we consider that 
$Y_f^{eff}$ originate in vacuum expectation values (VEVs)
of new scalars $Y_f$, i.e.
$$
Y_f^{eff} =\frac{y_f}{\Lambda} \langle Y_f\rangle ,
\eqno(1.2)
$$
where $\Lambda$ is a scale of the effective theory, i.e.
$$
H_Y =\sum_{i,j} \frac{y_u}{\Lambda} \, \overline{q}_L^i 
(Y_u)_i^j u_{Rj} H_u + \cdots .
\eqno(1.3)
$$
We refer the fields $Y_f$ as 
``Yukawaons" \cite{Yukawaon} 
hereafter.
Note that in the Yukawaon model, the Higgs scalars are
the same as ones in the conventional model, i.e. we
consider only two Higgs scalars $H_u$ and $H_d$ as the
origin of the masses (not as the origin of the
mass spectra).
It should be noted that 
the Yukawaons $Y_f$ are gauge singlets.

The VEVs of Yukawaons are obtained from supersymmetric
(SUSY) vacuum conditions for a superpotential $W$. 
In the conventional approach, the masses and mixings
are calculated by diagonalizing the mass matrices $M_f$
which are constrained by flavor symmetries, while,
in the Yukawaon approach, those are obtained
by writing a superpotential under flavor symmetries and 
by solving simultaneous equations from the SUSY vacuum 
conditions.
For example, we assume an O(3) flavor symmetry 
\cite{Koide-O3-PLB08} and we consider that the Yukawaons
$Y_f$ are $({\bf 3}\times{\bf 3})_S= {\bf 1}+{\bf 5}$ 
of O(3)$_F$:
$$
W_{Y}= \sum_{i,j} \frac{y_u}{\Lambda} u^c_i(Y_u)_{ij} {q}_{j} H_u  
+\sum_{i,j}\frac{y_d}{\Lambda} d^c_i(Y_d)_{ij} {q}_{j} H_d 
$$
$$
+\sum_{i,j} \frac{y_\nu}{\Lambda} \ell_i(Y_\nu)_{ij} \nu^c_{j} H_u  
+\sum_{i,j}\frac{y_e}{\Lambda} \ell_i(Y_e)_{ij} e^c_j H_d +h.c. 
+ \sum_{i,j}y_R \nu^c_i (Y_R)_{ij} \nu^c_j ,
\eqno(1.4)
$$ 
where $q$ and $\ell$ are SU(2)$_L$ doublet fields, and
$f^c$ ($f=u,d,e,\nu$) are SU(2)$_L$ singlet fields.
Here, in order to distinguish each $Y_f$ from others, 
we have assigned U(1)$_X$ charges as 
$Q_X(f^c)=-x_f$, $Q_X(Y_f)= +x_f$ and $Q_X(Y_R)=2x_\nu$.
We also write superpotential terms $W_f$ which are 
introduced in order to fix the VEVs of $Y_f$ under
the O(3) flavor symmetry and U(1)$_X$ symmetry,
and we obtain simultaneous equations for the VEVs 
$\langle Y_f\rangle$ by calculating SUSY vacuum 
conditions for $W=W_Y+W_u+W_d+ \cdots$.

In the next section, we give a short review of a mass 
spectrum of the charged leptons as an example of 
the supersymmetric Yukawaon approach.
We will discuss ratios
$$
R_e \equiv 
\frac{m_e +m_\mu +m_\tau}{(\sqrt{m_e} +\sqrt{m_\mu}
+\sqrt{m_\tau})^2} ,
\eqno(1.5)
$$
and
$$
r_e \equiv \frac{\sqrt{m_e m_\mu m_\tau}}{
(\sqrt{m_e} +\sqrt{m_\mu} +\sqrt{m_\tau} )^3} .
\eqno(1.6)
$$
In Sec.3, we propose a curious neutrino mass matrix
\footnote{
The form (1.7) has already been proposed in 
Ref.~3. Although five Yukawaons
$Y_u$, $Y_d$, $Y_e$, $Y_\nu$ and $Y_R$ have been
assumed in Ref.~3, 
in the present scenario, we will assume only 
four Yukawaons $Y_u$, $Y_d$, $Y_e$ and $Y_R$. 
}
$$
M_\nu = k_\nu M_e (M_e M_u^{1/2}+M_u^{1/2}M_e)^{-1} M_e,
\eqno(1.7)
$$
which is related to up-quark mass matrix $M_u$, and 
which can leads to a nearly tribimaximal mixing 
without assuming any discrete symmetry.
(The tribimaximal mixing has usually been explained by
assuming a discrete symmetry for the 
lepton mass matrices.\cite{tribi})
Finally, Sec.4 is devoted to summary and
concluding remarks.

\section{Mass spectrum}\label{sec2}
In order to see how to evaluate the mass spectra in the Yukawaon
model, let us show an example in the charged lepton sector.

Under the O(3) flavor symmetry and U(1)$_X$ symmetry, we can
write the following superpotential terms\cite{e-mass-Yukawaon}

$$
W_A= \lambda_A [\Phi_e \Phi_e A_e]+ 
\mu_A [Y_e A_e] 
+\lambda'_A [\Phi_e \Phi_e A'_e]+ 
\mu'_A [Y_e A'_e] 
$$
$$
+\lambda^{\prime\prime}_A [\hat{\Phi}_e \hat{\Phi}_e A'_e]+ 
\mu^{\prime\prime}_A  [Y'_e A'_e] ,
\eqno(2.1)
$$
where, for simplicity, we have denoted Tr$[X]$ as $[X]$ concisely.
Here, the field $\Phi_e$ has been introduced in order to fix 
the VEV values of $Y_e$.
We will refer $\Phi_e$ as an ``ur-Yukawaon".
The field $\hat{\Phi}_e$ denotes a traceless part of
the ur-Yukawaon $\Phi_e$, i.e. $\hat{\Phi}_e =\Phi_e -\frac{1}{3}[\Phi_e]$,
and we have assigned U(1)$_X$ charges as
$Q_X(Y_e)=x_e$, $Q_X(\Phi_e)=\frac{1}{2}x_e$ and  
$Q_X(A_e)=Q_X(A'_e)=-x_e$.
(In order to prevent $(Y'_e)_{ij}$ from coupling 
with $\ell_i e_j^c$, we have to assign a different U(1)$_X$ charge
to the field $Y'_e$ from the Yukawaon $Y_e$, so that we replace
\cite{e-mass-Yukawaon}
the coefficient $\mu^{\prime\prime}_A$ with 
$\lambda^{\prime\prime\prime}_A \phi_x$, where $Q_X(\phi_x)=x_\phi$ 
and $Q_X(Y'_e)=x_e -x_\phi$.
However, for simplicity, hereafter, we use the expression
$\mu^{\prime\prime}_A$ with U(1)$_X$ charge $x_\phi$ instead of 
$\lambda^{\prime\prime\prime}_A \phi_x$. )
In Eq.(2.1), we have assumed non-existence of a term
$[Y'_e A_e]$.
This is an ad hoc assumption, but it is a crucial assumption
to obtain a charged lepton mass relation (2.10) later.

From the SUSY vacuum condition $\partial W/\partial A_e=0$,  
and $\partial W/\partial A'_e=0$, 
we obtain the VEV relations 
$$
Y_e = k \Phi_e \Phi_e ,
\eqno(2.2)
$$
and 
$$
Y'_e  
= k' (\Phi_e \Phi_e +\xi \hat{\Phi}_e \hat{\Phi}_e) ,
\eqno(2.3)
$$
respectively, where $k=-\lambda_A/\mu_A$ and
$$
k' =-\frac{1}{\mu_A^{\prime\prime} }\left(
\lambda'_A - \frac{\mu'_A}{\mu_A} \lambda_A \right) , \ \ \
\xi= \frac{ \lambda_A^{\prime\prime} }{
\lambda'_A - \frac{\mu'_A}{\mu_A} \lambda_A } .
\eqno(2.4)
$$
Here, in Eq.(2.3), we have used the relation (2.2).

Next, we introduce a field $B_e$ with $Q_X=-\frac{3}{2}x_e +x_\phi$,
and we write a superpotential term
$$
W_B=\lambda_B [\Phi_e Y'_e B_e] +
\varepsilon_1 \lambda_B [\Phi_e][Y'_e][B_e] .
\eqno(2.5)
$$
(For a more general form, see Ref.\cite{e-mass-Yukawaon}.)
The SUSY vacuum condition $\partial W/\partial B_e =0$ 
($W=W_e=W_A+W_B$) gives 
$$
\Phi_e (\Phi_e\Phi_e +\xi \hat{\Phi}_e \hat{\Phi}_e) 
+\varepsilon_1 [\Phi_e] [\Phi_e\Phi_e +\xi \hat{\Phi_e} \hat{\Phi_e}]
{\bf 1}
$$
$$
=
(1+\xi)\Phi_e^3 - \frac{2}{3} \xi [\Phi_e]\, 
\Phi_e^2 + \frac{1}{9} \xi [\Phi_e]^2\,  \Phi_e 
$$
$$
+\varepsilon_1  [\Phi_e] \left( 
(1+\xi)[\Phi_e\Phi_e]-\frac{\xi}{3}[\Phi_e]^2 \right) 
{\bf 1} = 0 ,
\eqno(2.6)
$$
from Eq.(2.3).
(Other SUSY vacuum conditions $\partial W/\partial Y_e=0$
and $\partial W/\partial \Phi_e=0$ lead to $A_e=B_e=0$.)
On the other hand, in general, in a cubic equation
$$
\Phi^3 +c_2 \Phi^2 +c_1 \Phi + c_0 {\bf 1} = 0 ,
\eqno(2.7)
$$
the coefficients $c_i$ have the following relations:
$$
c_2 = -[\Phi] , \ \ c_1 =\frac{1}{2} \left(
[\Phi]^2 -[\Phi \Phi]\right) , 
\ \ c_0= - {\rm det}\Phi .
\eqno(2.8)
$$
In order to get non-zero a solution $[\Phi_e] \neq 0$,
we must take
$$
\xi= -3 ,
\eqno(2.9)
$$
from the coefficient $c_2$.
Then, we can obtain the ratios
$$
R_e \equiv  \frac{m_e +m_\mu + m_\tau}{
(\sqrt{m_e} +\sqrt{m_\mu} + \sqrt{m_\tau})^2 }
=\frac{[\Phi_e\Phi_e]}{[\Phi_e]^2}
= 1-\frac{2\xi}{9(1+\xi)}  = \frac{2}{3} ,
\eqno(2.10)
$$
and
$$
r_e =\frac{\sqrt{m_e m_\mu m_\tau}}{
(\sqrt{m_e} +\sqrt{m_\mu} +\sqrt{m_\tau})^3}
=\frac{ {\rm det}\Phi_e}{ [\Phi_e]^3} 
=- \frac{1}{6} \varepsilon_1 ,
\eqno(2.11)
$$
from the coefficients $c_1$ and $c_0$, respectively.
Thus, we can obtain the successful relation \cite{Koidemass}
(2.10) for the charged lepton masses.
At present, the parameter $\varepsilon_1$ is free, so that 
we cannot predict the value of $r_e$.

By the way, since we have successfully obtained the relation (2.10)
without any adjustable parameter,
another problem has risen in the present scenario:
We know that $R=2/3$ is valid only for the charged
lepton masses, and the observed masses for
another sectors do not satisfy $R=2/3$. 
For example, the ratio $R_u$ for the up-quark masses
is $R_u \simeq 8/9$ \cite{Koide-JPG07}.
Can we modify the present scenario as it leads to
$R_u\simeq 8/9$?
At present it seems to be impossible, because there is
no adjustable parameter in the present scenario. 
We expect that all Yukawaons are related to each other and
those VEVs are described in 
terms of the VEV of the ur-Yukawaon $\Phi_e$. 
(A possibility that $\langle Y_u\rangle$ is described
in terms of $\langle Y_e\rangle$ is discussed in 
Ref.~\cite{e-mass-Yukawaon} ).
However, the purpose of the present talk is not to 
give the mass spectra, we do not discuss more details
of this topic here.
Our goal is to give a unified description of all 
Yukawaons.
As the first step, we discuss a phenomenological 
relation between the Yukawaons $Y_e$ and $Y_u$
in the next section.

\section{Neutrino mass matrix}
Based on a supersymmetric Yukawaon model, a curious neutrino 
mass matrix has recently been proposed.\cite{Koide-O3-PLB08}.
The mass matrix $M_\nu$ is related to up-quark masses as follows:
$$
M_\nu = M_D M_R^{-1} M_D^T ,
\eqno(3.1)
$$
where the neutrino Dirac mass matrix $M_D$ is given by 
$$
M_D \propto \langle Y_\nu\rangle \propto 
\langle Y_e\rangle \propto M_e ,
\eqno(3.2)
$$
and the right-handed neutrino Majorana mass matrix $M_R$ is given by
$$
M_R \propto \langle Y_R \rangle \propto  
\langle Y_e\rangle \langle \Phi_u \rangle 
+ \langle \Phi_u \rangle \langle Y_e \rangle
 \propto M_e M_u^{1/2} + M_u^{1/2} M_e ,
\eqno(3.3)
$$
where the ur-Yukawaon $\Phi_u$ has a relation 
$Y_u =k_u \Phi_u \Phi_u$ similar to Eq.(2.2).

In the present model, differently from the model given in 
Ref.~\cite{Koide-O3-PLB08}, we assume that $\nu^c$ and
$e^c$ have the same U(1)$_X$ charge $x_e$, so that 
the Yukawaon $Y_e$ couples not only to the charged lepton
sector, but also to the neutrino sector:
$$
W_{Y}= 
 \frac{y_\nu}{\Lambda} (\ell Y_e \nu^c) H_u  
+\frac{y_e}{\Lambda} (\ell Y_e e^c) H_d 
+ y_R (\nu^c Y_R \nu^c) 
+ \frac{y'_R}{\Lambda} (\nu^c Y_e Y_e \nu^c) .
\eqno(3.4)
$$
Next, we assume additional fields $A_R$ with 
$Q_X=-(\frac{1}{2}x_u + x_e)$, so that we obtain 
the superpotential terms for $A_R$ as follows:\footnote{
Exactly speaking, we have to read $\mu_R$ as $\lambda'_R \phi_R$.
Otherwise, the field $Y_R$ has the U(1)$_X$ charge
$Q_X(Y_R)=2 x_e =\frac{1}{2}x_u + x_e$, so that we are
obliged to accept the relation $x_e=\frac{1}{2}x_u$.
This means that $Y_e$ and $\Phi_u$ have the same charge,
a mixing between $Y_e$ and $\Phi_u$ is caused. 
}
$$
W_R = \lambda_{R} [(Y_e \Phi_u +\Phi_u Y_e) A_R]
+ \mu_R [Y_R A_R] .
\eqno(3.5)
$$
Then, from the SUSY vacuum condition $\partial W/\partial A_R=0$,
we can obtain the relation (3.3).

In order to calculate lepton mixing matrix $U$,
we have to know a matrix form  $M_\nu$ at the diagonal basis of 
$\langle Y_e\rangle$.  
Hereafter, we will denote a VEV matrix of a field $A$ 
at the diagonal basis of $\langle Y_f\rangle$ as 
$\langle A\rangle_f$.
From the definition of the diagonal basis, we can express
$$
\langle Y_e \rangle_e \propto D_e = {\rm diag}(m_e, m_\mu, m_\tau),
\eqno(3.6)
$$
$$
 \langle \Phi_u \rangle_e \neq \langle \Phi_u \rangle_u
\propto D_u^{1/2}={\rm diag}\left( \sqrt{m_u}, \sqrt{m_c},
\sqrt{m_t}\right) .
\eqno(3.7)
$$
Our concern is in the form of $\langle \Phi_u \rangle_e$.
The Cabibbo-Kobayashi-Maskawa (CKM) matrix $V$ satisfies 
the following relation\footnote{
In the present O(3) flavor model, the mass matrix $M_f$ is
diagonalized as $U_f^T M_f U_f = D_f$.
}
$$
\langle Y_u \rangle_d = V^T (\delta_q)\langle Y_u \rangle_u 
V(\delta_q) .
\eqno(3.8)
$$
On the analogy of the form of $\langle Y_u \rangle_d$, 
(3.8),  
we assume a form of $\langle \Phi_u \rangle_e$ as 
$$
\langle \Phi_u \rangle_e = V^T (\delta_\ell)\langle \Phi_u 
\rangle_u V(\delta_\ell) ,
\eqno(3.9)
$$
where we have assumed that $\delta_\ell$ is a free parameter.
Then, from observed values of the up-quark masses\cite{q-mass} 
and CKM parameters\cite{PDG08}, we obtain numerical results 
of the neutrino mixing parameters $\sin^2 2\theta_{23}$, 
$\tan^2 \theta_{12}$ and $|U_{13}|$ 
as shown in Table 1.

\begin{table}
\begin{center}
{Table 1. Numerical results for neutrino mixing parameters 
versus $\delta_\ell$.}

\vspace{2mm}

{\begin{tabular}{@{}clllll@{}} \hline
$\delta_\ell$ &  $\sin^2 2\theta_{23}$ & $\tan^2 \theta_{12}$ &
$|U_{13}|$ & $\Delta m^2_{21}/\Delta m^2_{32}$ \\ \hline
0    &  $0.4803$ & $0.4745$ & $0.01042$ & $0.00196$ \\
$60^\circ$ & $0.7631$ & $0.4801$ & $0.00844$ & $0.00139$\\
$68.9^\circ$ & $0.8127$ & $0.4851$ &$0.00781$ & $0.00127$ \\
$90^\circ$ & $0.9028$ & $0.5017 $ & $0.00615$ & $0.00102$ \\
$120^\circ$ & $0.9688$ & $0.5277$ & $0.00386$ & $0.00081$ \\
 $180^\circ$ &  $0.9952$ & $0.5525$ & $0.00094$ & $0.00068$ \\
\hline 
\end{tabular}
}
\end{center}
\end{table}

The observed value\cite{PDG08} of $\delta_q$ in the quark 
sector is $\delta_q=68.9^\circ$, but the value 
$\delta_\ell=\delta_q=68.9^\circ$
can not give a reasonable value of $\sin^2 2\theta_{23}$.
On the other hand, the value $\delta_\ell=180^\circ$ can 
successfully give a nearly tribimaximal mixing. 
Since, in the present O(3) model, the parameter       
$\delta_\ell$ must be $0$ or $\pi$ because we have 
assumed that $\langle\Phi_e\rangle$ and 
$\langle\Phi_u\rangle$ are real.
This is in favor of the results in Table 1. 
Although the values of $\Delta m^2_{21}/\Delta m^2_{32}$
in Table 1 are too small compared with the observed
value\cite{MINOS,KamLAND} 
$\Delta m^2_{21}/\Delta m^2_{32}=0.028\pm 0.004$,
the values can suitably be adjusted by considering the
$y'_R$-term in Eq.(3.4) without changing the
predictions of the neutrino mixing parameters.

\section{Concluding remarks}

In conclusion, we have obtained a curious neutrino
mass matrix (1.7) which is related to up-quark mass matrix, 
and which can leads to a nearly tribimaximal mixing
without assuming any discrete symmetry.
(Although the form (1.7) has already been given in 
Ref.~\cite{Koide-O3-PLB08}, the superpotential form
in the present Yukawaon model is somewhat different from
that in Ref.~\cite{Koide-O3-PLB08}.)
However, at present, there is no theoretical ground 
for the relation (3.9).
The relation (3.9) is merely a phenomenological
assumption.
Nevertheless, we have successfully obtained a nearly 
tribimaximal mixing under this assumption without 
assuming any discrete symmetry.
It will offer an important clue to a unified 
understanding of quarks and leptons to investigate
why the phenomenological assumption (3.9) is so 
effectual.

The approach based on a Yukawaon model to
masses and mixings will provide a new view
different from conventional mass matrix models.
Especially, the approach seems to be powerful to the
predictions of mass relations $R_f$ and $r_f$
similar to $R_e$ and $r_e$ defined in Eqs.(1.5) and
(1.6), respectively.

The neutrino mass matrix (1.7) can successfully give 
a tribimaximal mixing without assuming any discrete symmetry.
Here, the neutrino mass matrix has been described by
VEVs of the ur-Yukawaons $\Phi_e$ and $\Phi_u$.
This suggests a possible relation between Yukawaons
in the lepton and quark sectors. 
Our goal is to give a unified description of all 
Yukawaons.
There is a possibility that VEVs of all Yukawaons $Y_f$ are
described only one VEV matrix of the ur-Yukawaon $Y_e$. 
So far, we did not discuss the down-quark Yukawaon $Y_d$.
Whether a unified Yukawaon model is possible or not
is dependent on whether a $Y_d$ can also reasonably be
described in terms of $\Phi_e$ (and also $\Phi_u$).
This will be a touchstone of the Yukawaon approach.

At present, the calculations have been done in the
supersymmetric limit. 
We have considered  
$\langle \Phi_f \rangle \sim \Lambda \sim 10^{15}$ GeV,
so that almost Yukawaon components are massive and invisible 
in the low energy phenomena. 
However, some of Yukawaons are massless in the 
supersymmetric limit. 
Since we consider an explicitly broken term of the
flavor symmetry which effectively appears when SUSY is 
broken, those massless particles will acquire masses of
TeV scale, so that the effects will 
be able to be observed in the TeV region physics.


\end{document}